\documentclass[final]{wl-ws}

\def\privnote#1{}
\newcommand{\Operator}[1]{\mathchoice
   {\mbox{\boldmath $#1$}}{\mbox{\boldmath $#1$}}
   {\mbox{\footnotesize \boldmath $#1$}}
   {\mbox{\footnotesize \boldmath $#1$}}} %kein \boldmath in \scriptsize
\def\aside#1{(#1)}

\def\lstop#1{\left.}
\def\rstop#1{\right.}
\def\bbbr{{\rm I\!R}}
\font \fivesans               = cmss10 at 5pt
\font \sevensans              = cmss10 at 7pt
\font \tensans                = cmss10
\newfam\sansfam
\textfont\sansfam=\tensans\scriptfont\sansfam=\sevensans
\scriptscriptfont\sansfam=\fivesans
\def\sans{\fam\sansfam\tensans}
\def\bbbz{{\mathchoice {\hbox{$\sans\textstyle Z\kern-0.4em Z$}}
{\hbox{$\sans\textstyle Z\kern-0.4em Z$}}
{\hbox{$\sans\scriptstyle Z\kern-0.3em Z$}}
{\hbox{$\sans\scriptscriptstyle Z\kern-0.2em Z$}}}}
\def\testspace#1{\mbox{$\cal #1$}}
\def\stackscript#1{\mathop{\smash{#1}\vphantom{#1}}\limits}
\newcommand{\region}[1]{\mbox{$\cal #1$}}

\def\sqr#1#2{{\vcenter{\hrule height.#2pt
    \hbox{\vrule width.#2pt height#1pt \kern#1pt
      \vrule width .#2pt}
    \hrule height.#2pt}}}

\def\Op#1{{\Operator #1}}
\def\norm#1{\left\| #1 \right\|}
\def\modulus#1{\left| #1 \right|}
\def\xv{{\vec x}}
\def\Nv{{\vec \nabla}}

\input prepictex \input pictex \input postpictex

\begin{document}

\title{\phantom .\vskip -25mm \hfill{\normalsize
ASI-TPA/1/99}\linebreak\vskip 1mm Nonlocality in Nonlinear Quantum
Mechanics\thanks{Based on a talk presented at ``New Insights in Quantum
Mechanics -- Fundamentals, Experimental Results, Theoretical Directions'',
Goslar, August 31 -- September 3, 1998.}}

\author{W. L\"ucke}

\address{Arnold Sommerfeld Institute for Mathematical Physics, Technical
University Clausthal, Leibnizstr.~10, D-38678 Clausthal,
GERMANY\\E-mail: aswl@pt.tu-clausthal.de}

\maketitle

\abstracts{A general method for testing essential nonlocality of
nonlinear modifications of quantum mechanics is presented and applied to
show the inconsistency of I.~Bialynicki-Birula's and
J.~Mycielski's nonlinear quantum theory.}

\section{Nonlinear Modifications of Quantum Mechanics}

Many authors\footnote{Let us mention just some of them:
T.W.B.~Kibble\,\cite{Kibble}, A.~Ashtekar and
T.A.~Schilling\,\cite{Ashtekar}, P.~Bona\,\cite{Bona}, R.~Haag and
U.~Bannier\,\cite{Haag-Bannier}, Mielnik\,\cite{Mielnik},
S.~Weinberg\,\cite{Weinberg}, G.~Auberson and
P.C.~Sabatier\,\cite{Sabatier}, M.D.~Kostin\,\cite{Kostin}, H.D.~Doebner
and Goldin\,\cite{dg1,dg2}, I.~Bialynicki-Birula and
J.~Mycielski\,\cite{bbm}.} considered nonlinear Schr\"odinger equations of
the form
\begin{equation} \label{NLSE}
i\hbar\frac{\partial}{\partial t} \Psi_t(\xv) =
\left(-\frac{\hbar^2}{2m}\Delta+V(\xv,t)\right) \Psi_t(\xv) +
R[\Psi_t](\xv) \Psi_t(\xv)\,,
\end{equation}
where $F[\Psi] = R[\Psi] \Psi$ is a local\footnote{Here locality means
$\Psi(\xv)=0\Longrightarrow \left(F[\Psi]\right)(\xv) =0$ in the
distribution theoretical sense.} nonlinear Functional of $\Psi\,$.
The essential point is that (\ref{NLSE}) is not interpreted as an
equation for some field operator but as a classical evolution equation
for the quantum mechanical wave function:
\begin{equation} \label{StInt}
\int_{{\cal B}} \modulus{\Psi_t(\xv)}^2\,{\rm d}\xv =
\parbox{6cm}{probability for location within ${\cal \region B}$}
\end{equation}
(at time $t$ for normalized $\Psi_t$).
Special cases were even tested experimentally.\,\cite{tests}

\vskip 5mm

\noindent
All these efforts seemed useless according to N.~Gisin's
claim\,\cite{gis1,gis2}:
\begin{equation} \label{gis-cl}
\mbox{``All deterministic nonlinear Schr\"odinger equations are irrelevant.''}
\end{equation}
By this Gisin meant the following:
\begin{quote}
Consider a Bell-like situation as sketched in Figure \ref{fig:gisin}.
\begin{figure}[ht]
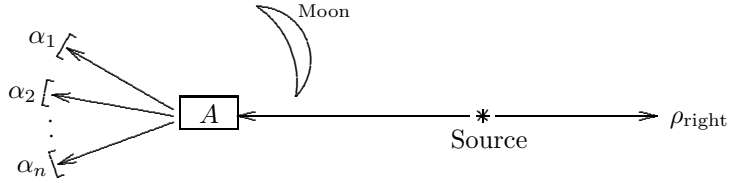
%[t]
$$
\font\thinlinefont=cmr5
\mbox{\beginpicture
\setcoordinatesystem units <0.9cm,0.9cm>
\unitlength=1.04987cm
\linethickness=1pt
\setplotsymbol ({\makebox(0,0)[l]{\tencirc\symbol{'160}}})
\setshadesymbol ({\thinlinefont .})
\setlinear
%
% Fig CIRCULAR ARC object
%
\linethickness= 0.500pt
\setplotsymbol ({\thinlinefont .})
\circulararc 134.333 degrees from  4.572 24.575 center at  4.006 25.121
%
% Fig CIRCULAR ARC object
%
\linethickness= 0.500pt
\setplotsymbol ({\thinlinefont .})
\circulararc 65.361 degrees from  4.572 24.575 center at  3.247 24.796
%
% Fig ELLIPSE
%
\linethickness= 0.500pt
\setplotsymbol ({\thinlinefont .})
\put{\makebox(0,0)[l]{\circle*{ 0.047}}} at  0.948 24.213
%
% Fig ELLIPSE
%
\linethickness= 0.500pt
\setplotsymbol ({\thinlinefont .})
\put{\makebox(0,0)[l]{\circle*{ 0.047}}} at  0.961 23.995
%
% Fig POLYLINE object
%
\linethickness= 0.500pt
\setplotsymbol ({\thinlinefont .})
\putrule from  7.144 24.289 to  3.715 24.289
%
% arrow head
%
\plot  3.969 24.352  3.715 24.289  3.969 24.225 /
%
%
% Fig POLYLINE object
%
\linethickness= 0.500pt
\setplotsymbol ({\thinlinefont .})
\putrectangle corners at  2.857 24.575 and  3.715 24.098
%
% Fig POLYLINE object
%
\linethickness= 0.500pt
\setplotsymbol ({\thinlinefont .})
\plot  2.762 24.384  1.194 25.290 /
%
% arrow head
%
\plot  1.446 25.218  1.194 25.290  1.382 25.108 /
%
%
% Fig POLYLINE object
%
\linethickness= 0.500pt
\setplotsymbol ({\thinlinefont .})
\plot  1.321 25.453  1.238 25.502 /
\plot  1.238 25.502  1.048 25.171 /
\plot  1.048 25.171  1.130 25.125 /
%
% Fig POLYLINE object
%
\linethickness= 0.500pt
\setplotsymbol ({\thinlinefont .})
\plot  2.762 24.289  0.978 24.604 /
%
% arrow head
%
\plot  1.239 24.622  0.978 24.604  1.217 24.497 /
%
%
% Fig POLYLINE object
%
\linethickness= 0.500pt
\setplotsymbol ({\thinlinefont .})
\plot  0.991 24.807  0.897 24.824 /
\plot  0.897 24.824  0.830 24.448 /
\plot  0.830 24.448  0.923 24.431 /
%
% Fig POLYLINE object
%
\linethickness= 0.500pt
\setplotsymbol ({\thinlinefont .})
\plot  2.766 24.196  1.065 23.575 /
%
% arrow head
%
\plot  1.282 23.722  1.065 23.575  1.325 23.603 /
%
%
% Fig POLYLINE object
%
\linethickness= 0.500pt
\setplotsymbol ({\thinlinefont .})
\plot  1.016 23.779  0.929 23.747 /
\plot  0.929 23.747  1.056 23.387 /
\plot  1.056 23.387  1.145 23.421 /
%
% Fig POLYLINE object
%
\linethickness= 0.500pt
\setplotsymbol ({\thinlinefont .})
\putrule from  7.239 24.289 to  7.429 24.289
%
% Fig POLYLINE object
%
\linethickness= 0.500pt
\setplotsymbol ({\thinlinefont .})
\putrule from  7.334 24.384 to  7.334 24.194
%
% Fig POLYLINE object
%
\linethickness= 0.500pt
\setplotsymbol ({\thinlinefont .})
\plot  7.262 24.361  7.398 24.225 /
%
% Fig POLYLINE object
%
\linethickness= 0.500pt
\setplotsymbol ({\thinlinefont .})
\plot  7.404 24.369  7.269 24.234 /
%
% Fig POLYLINE object
%
\linethickness= 0.500pt
\setplotsymbol ({\thinlinefont .})
\putrule from  7.525 24.289 to  9.906 24.289
%
% arrow head
%
\plot  9.652 24.225  9.906 24.289  9.652 24.352 /
%
%
% Fig TEXT object
%
\put{$A$} [lB] at  3.143 24.194
%
% Fig TEXT object
%
\put{$\alpha_n$} [lB] at  0.45 23.461
%
% Fig TEXT object
%
\put{$\alpha_1$} [lB] at  0.622 25.364
%
% Fig TEXT object
%
\put{$\alpha_2$} [lB] at  0.32 24.558
%
% Fig TEXT object
%
\put{$\rho_{\rm right}$} [lB] at 10.097 24.194
%
% Fig TEXT object
%
\put{Source} [lB] at  6.858 23.812
\put{\scriptsize Moon} [lB] at  4.6 25.75
\linethickness=0pt
\putrectangle corners at  0.434 25.929 and 10.097 23.362
\endpicture}
$$
\caption{Gisin's Gedanken experiment. \label{fig:gisin}}
\end{figure}
Then, if the source produces entangled 2-particle states, there is
always a physical observable for the particle sent to the
right,\footnote{Note that, contrary to Bell nonlocality,\cite{cl-sh}
the observable effect does not refer to the resulting correlations
between both particles.} the probability distribution of which is
instantaneously (substantially) changed by suitable measurements
(involving only low energy transfer) on the other particle `behind the
moon'.
\end{quote}

\vskip 5mm

\noindent
Actually, Gisin\,\cite{gis1} assumed the following:\footnote{His
justification: ``So far we have only used linear quantum mechanics''.}
\begin{quote}
\begin{itemize}
\item[(G1):]
If the observable $A$ resp.~$B$ of the particle `behind the moon' is
measured at time $t=0$ the partial state of the other particle at times
$t\geq0$ is given by a density matrix $\rho_{\rm right}(t)$ of the
form\footnote{This especially implies $\sum_{\alpha} x_\alpha\,
P_\Psi(\alpha,t) = \sum_{\beta} x_\beta\, P_\Psi(\beta,t)$ for $t=0$ but
--- as realized by Gisin --- not generally for $t>0\,$.}
$$
\sum_{\alpha} x_\alpha\, P_\Psi(\alpha,t)\quad
\mbox{resp.}\quad \sum_{\beta} x_\beta\,
P_\Psi(\beta,t)\,,
$$
where the $P_\Psi(\alpha,t)$ resp.~$P_\Psi(\beta,t)$ are pure states 
evolving according to the corresponding 1-particle equation.
\item[(G2):]
{\bf All} self-adjoint \aside{bounded} operators correspond to observables.
\end{itemize}
\end{quote}

\vskip 5mm

That Gisin's claim (\ref{gis-cl}) is wrong since Gisin's assumption
(G1), the projection postulate, is unjustified in nonlinear quantum
mechanics has already been pointed out by Polchinski\,\cite{pol}, who
determined conditions which are sufficient for the absence of essential
nonlocality.\footnote{See also \cite{czach1} for examples.}

\vskip 5mm

Accepting Gisin's assumption (G2), Polchinski concluded that his
conditions -- violated for prominent examples of nonlinear Schr\"odinger
equations -- are also necessary to avoid essential nonlocality. However,
as explained already in \cite{lu1}, also assumption (G2) is unjustified
in nonlinear quantum mechanics and definitely wrong for the situation
reconsidered in the next section. This is why valid proofs of essential
nonlocality have to be more involved.

\medskip

\section{What we can Learn from Nonlinear Gauge Transformations}
\label{section:CE}

Consider the well-defined special case
$$
\left(\Op N_{\! D}(\Psi)\right)(\xv) \stackrel{\rm def}{=}
e^{i\frac{2mD}{\hbar}\ln\modulus{\Psi(\xv)}}\Psi(\xv)\;,\quad D\in\bbbr\,.
$$
of the `nonlinear gauge transformations' exploited by H.-D.~Doebner et
al.\,\cite{nlgt} If $\Psi_t'(\xv)$ is a solution of (\ref{NLSE}) for $R=0$ then
straightforward calculation shows that
$$
\Psi_t(\xv) = \left(\Op N_{\! D}(\Psi'_t)\right)(\xv)
$$
is a solution of (\ref{NLSE}) for
\begin{equation} \label{GDGE}
R[\Psi_t] = \hbar D \Bigl(\frac{i}{2}
\frac{\Delta\rho_t}{\rho_t}\Psi_t + c_1 \frac{\Nv\cdot \vec
{J}_t}{\rho_t} + c_2 \frac{\Delta\rho_t}{\rho_t} + c_3
\frac{\vec{J}_t}{\rho_t}+ +c_4 \frac{\vec{J}_t\cdot
\Nv\rho_t}{(\rho_t)^2} + c_5 \frac{(\Nv\rho_t)^2}{(\rho_t)^2}\Bigr)
\end{equation}
with\footnote{The {\em general Doebner-Goldin equation\/} is given by
(\ref{NLSE}) and (\ref{GDGE}) without the restriction (\ref{el}) on the
parameters $c_\nu\in\bbbr\,$.}
\begin{equation} \label{el}
c_1=1\;, \quad c_2=-2c_5=-mD/\hbar\;,\quad c_3=0\;,\quad c_4=-1\,,
\end{equation}
where
$$
\rho_t\stackrel{\rm def}{=}\modulus{\Psi_t}^2\,,\; \vec{J}_t
\stackrel{\rm def}{=} \frac{1}{2i} \left( \overline{\Psi_t} \Nv \Psi_t
- \Psi_t \Nv \overline{\Psi_t} \right)\,.
$$
This way we get a deterministic nonlinear Schr\"odinger equation which,
interpreted by (\ref{StInt}), describes the same physics as the
corresponding ($R=0$) linear Schr\"odinger equation, since
$$
\modulus{\Psi_t(\xv)} = \modulus{\Psi'_t(\xv)}\,.
$$
That, contrary to Gisin's as well as Polchinski's claim, there is no
real problem with locality is no surprise since now nonlinear
projection operators
$$
\Op E = \Op N_D \circ \Op E' \circ \Op N_D^{-1}\,,
$$
instead of linear projectors $\Op E'\,$, have to be used\,\cite{lu1} to
get the correct probabilities
$$
\norm{\Op E(\Psi_t)}^2 = \norm{\Op E'(\Psi'_t)}^2 = \left\langle
\Psi'_t \mid \Op E'\,\Psi'_t \right\rangle\,.
$$
Hence assumption (G2) is obviously wrong in nonlinear quantum mechanics.
Moreover, w.r.t.~the nonlinear $\Op E\,$, density matrices are
inadequate for the description of classical mixtures:
$$
\sum_{\alpha} x_\alpha \norm{\Op E(\Psi_\alpha)}^2 = {\rm
trace}\left(\Op E'\, \sum_\alpha x_\alpha \Op P_{\Psi'_\alpha}\right)
\stackrel{\rm i.g.}{\ne} {\rm trace}\left(\Op E\, \sum_\alpha x_\alpha
\Op P_{\Psi_\alpha}\right)\,.
$$
Therefore also assumption (G1) turns out to be quite inadequate for
nonlinear modifications of quantum mechanics.

The simple example (\ref{NLSE})/(\ref{GDGE})/(\ref{el}) tells us that
essential nonlocality should to be checked by using nothing else than
the evolution equation together with its basic interpretation
(\ref{StInt}).

\medskip

\section{The Doebner-Goldin Equation Interpreted as 2-Particle Equation}
\label{section:dg}

Let us interpret (\ref{NLSE}) as a two-particle equation:
$$
\xv=(\xv_1,\xv_2)\;,\quad \xv_j = \mbox{position of particle } j\,.
$$
Moreover assume that the potential is of the form
$$
V(\xv) = V_2(\xv_2-\xv_0)\;,\quad \xv_0 \mbox{ fixed}\,,
$$
and that $\Psi_t(\xv)$ is the solution of (\ref{NLSE}) fulfilling the
initial condition
$$
\Psi_0(\xv_1,\xv_2) = f(\xv_1,\xv_2-\xv_0)\,.
$$
Then, obviously, we have an unacceptable nonlocality, if the position
probability density\footnote{Note that (\ref{pp1}) does not depend on
$\xv_0\,$. Therefore the effect on particle 1, if any, can be produced by
acting on particle 2 as far away as one likes.}
\begin{equation} \label{pp1}
\rho_1(\xv_1,t) \stackrel{\rm def}{=}
\int \modulus{\Psi_t(\xv_1,\xv_2)}^2{\rm d}\xv_2
\end{equation}
for particle 1 depends on $V_2\,$.

That the latter happens for certain cases of the general Doebner-Goldin
equation, if interpreted as 2-particle equation, was first proved by
Werner.\,\cite{werner1} Inspired by E.~Nelson\,\cite{nelson} he
considered pairs of 1-dimensional particles,
i.e.~$\xv=(x_1,x_2)\in\bbbr^2\,$, with
\begin{equation} \label{osz}
V_2(x_2) = \lambda\,(x_2)^2\;,\quad \lambda\in\bbbr\,,
\end{equation}
and entangled initial conditions of Gaussian type. The corresponding
solutions are of the form
$$
\Psi_t(x_1,x_2)= e^{\gamma(t)
-\sum_{j,k=1}^2 C_{jk}(t)x_jx_k/2}\;,\quad C_{jk}=C_{kj}\,,
$$
where the $C_{jk}(t)$ fulfill a simple system of first order ordinary
differential equations that can be used to determine their time
derivatives at $t=0$ and thus
\begin{equation} \label{test1}
\left({\partial_t}^n\int (x_1)^2 \rho_1(x_1,t)\,{\rm
d}x_1\right)_{|_{t=0}}
\end{equation}
for arbitrary $n\in\bbbz_+\,$. Werner found that (\ref{test1}) depends on
$\lambda$ for $n=3$ and suitable $C_{jk}(0)$ unless
\begin{equation} \label{gi}
c_3=c_1+c_4=0\,.
\end{equation}
In principle, using (\ref{NLSE}) directly,\,\cite{luna} one may calculate
\begin{equation} \label{test2}
\partial_\lambda \left(\partial_t^n \int\modulus{\Psi_t(x,y)}^2\,{\rm
d}y\right)_{|_{t=0}}
\end{equation}
as a functional of $\Psi_0$ and $V$ for given $R\,$. This shows that
(\ref{test1}) varies also with strictly localized changes of $V_2\,$.
However, for $n>3$ the calculation becomes too involved and could not
even managed by use of computer algebra. On the other hand, Werner's
Ansatz turned out to be too special to uncover essential nonlocality of
the Galilei covariant cases of the Doebner-Goldin 2-particle
equation.\footnote{Note that (\ref{gi}) is equivalent to Galilei
covariance of the Doebner Goldin equation.\,\cite{dg2}} This is why
\begin{equation} \label{test3}
\left({\partial_t}^n\int x_1 \rho_1(x_1,t)\,{\rm
d}x_1\right)_{|_{t=0}}
\end{equation}
was checked for $n=4$ in \cite{lu2} showing that the Doebner-Goldin
equation is essentially nonlocal, when interpreted as a 2-particle
equation, unless the parameters $c_\nu$ are chosen such that
(\ref{NLSE}) is (formally) linearizable by some nonlinear gauge
transformation.

\medskip

\section{Inconsistency of Bialynicki-Birula's and Mycielski's Theory}
\label{section:bbm}

For the nonlinear Schr\"odinger equation of Bialynicki-Birula and
Mycielski, given by\footnote{Equation (\ref{bbm}) was already
considered by H.~Ko\v s\v t\'al\,\cite{kostal} with $b<0\,$.}
\begin{equation} \label{bbm}
\left(R[\Psi]\right)(\xv) = -2b\,\ln\modulus{\Psi(\xv)}\;,\quad b\in\bbbr\,,
\end{equation}
testing (\ref{test3}) is of no use, since the Ehrenfest relations
hold.\,\cite{bbm} In such cases one should check
\begin{equation} \label{test4}
S_{k,n}[\Psi_0,V_2] \stackrel{\rm def}{=} \partial_\lambda
\left(\partial_t^n \int e^{ikx}\,\modulus{\Psi_t(x,y)}^2 {\rm d}x\,{\rm
d}y\right)_{|_{t=0}}
\end{equation}
for $k\in\bbbr$ and $n\in\bbbz_+\,$. In principle this is equivalent to
testing (\ref{test2}) but has the advantage of testing (\ref{test3}):
The integral over $x$ allows for partial integrations which simplify the
resulting expressions considerably.

\vskip 5mm

\noindent
For simplicity, let $\frac{\hbar^2}{2m}=1$ and assume $R$ to be real
valued,\footnote{For the Doebner-Goldin case (\ref{GDGE}) the latter can
always be achieved\,\cite{NaDip} by some nonlinear gauge transformation
of the type considered in Section \ref{section:CE}.} as in (\ref{bbm}):
\begin{equation} \label{rR}
\left(R[\Psi]\right)(\xv) = \overline{\left(R[\Psi]\right)(\xv)}\,.
\end{equation}
Then, defining
$$
\begin{array}[c]{rcl}
T_{k,\nu}(t) &\stackrel{\rm def}{=}&\displaystyle  \int
e^{ikx}\,\overline{\Psi_t}\,\partial_x^\nu \Psi_t\,{\rm d}x\,{\rm d}y\,,\\
\displaystyle D_{k,\mu,\nu}(t) &\stackrel{\rm def}{=}&\displaystyle 
\int e^{ikx}\,\left(\partial_x^\mu R[\Psi_t]\right)\overline{\Psi_t}\,
\partial_x^\nu \Psi_t\,{\rm d}x\,{\rm d}y\,,
\end{array}
$$
we get from (\ref{NLSE}) by partial integration
$$
i\frac{{\rm d}}{{\rm d} t} T_{k,\nu}= -k^2\, T_{k,\nu} + 2ik\, T_{k,\nu+1}
+ \sum_{\mu=1}^{\nu}{\nu\choose \mu} D_{k,\mu,\nu-\mu}
$$
for $\xv=(x_1,x_2)\in\bbbr^2$ and $V(\xv)=V_2(x_2)\,$.
Iteration of this gives
$$
\begin{array}[c]{rcl}
\displaystyle \left(i \frac{{\rm d}}{{\rm d} t}\right)^3 T_{k,0}
&=& -k^6\, T_{k,0} + 6ik^5\, T_{k,1} + 12k^4 T_{k,2} - 8 ik^3 T_{k,3}\\
&& -\left(4ik^3\,D_{k,1,0} +8k^2 D_{k,1,1} + 4k^2 D_{k,2,0} +
2k\partial_t D_{k,1,0}\right)\,.
\end{array}
$$
Hence, e.g.,
$$
\begin{array}[c]{rcl}
\displaystyle \left(\partial_t^{n+3}\!\int
x^2\modulus{\Psi_t(x,y)}^2{\rm d}x\,{\rm d}y \right)_{|_{t=0}}
&=&\displaystyle  -\left(\partial_t^{n+3}\,(\partial_k)^2
T_{k,0}(t)\right)_{|_{t=k=0}}\\[4mm]
&=& 8i\,\partial_t^n\left(2\,D_{0,1,1} + D_{0,2,0}\right).
\end{array}
$$
For the special case
$$
m=1\;,\quad V_2(y) = \lambda\,y^2\;,\quad \Psi_0(x,y) = \frac{e^{-x^2 -y^2
-xy}}{\int e^{-2x^2 -2y^2  -2xy}\, {\rm d}x\,{\rm d}y}\,,
$$
running a simple computer algebra program (see appendix) shows that
(\ref{NLSE}) and (\ref{bbm}) imply\footnote{This result was confirmed by
R.~Werner\,\cite{werner2} using his method described in Section
\ref{section:dg}.}
$$
\partial_\lambda\left((i\partial_t)^3\left(2\,D_{0,1,1}(t) +
D_{0,2,0}(t)\right)\right)_{|_{t=0}} = 32b\,.
$$
This means that (\ref{NLSE})/(\ref{bbm}) is essential nonlocal ---
against the basic philosophy of Bialynicki-Birula's and J.~Mycielski's
theory.\cite{bbm}

\medskip

\section{Identical Particles} 

Up to now we tacitly assumed that the two particles (with equal masses)
considered in Figure \ref{fig:gisin} can be distinguished. Therefore one
might still hope that Bialynicki-Birula's and J.~Mycielski's theory is
consistent for identical particles. However, even for 2-particles states
which are symmetric or antisymmetric w.r.t.~exchange of the particles
essential nonlocality is unavoidable. To show this denote by
$\Psi_t^{g,U}$ the solution of
$$
i\partial_t \Psi_t^{g,U} = \left(-\Delta +\lambda U - 2b
\ln\modulus{\Psi_t^{g,U}}\right)\Psi_t^{g,U}
$$
fulfilling the initial condition
$$
\Psi_0^{g,U}(x,y) = g(x,y)\,.
$$
For fixed $f(x,y)$ and $\sigma\in\left\{+1,-1\right\}$ define
$$
\begin{array}[c]{c}
\begin{array}[c]{rcl}
U^{(d)}(x,y) &\stackrel{\rm def}{=}& V(y-d) + V(x-d)\,,\\
\Psi_0^{(d)}(x,y) &\stackrel{\rm def}{=}& f(x,y-d) +\sigma f(y,x-d)\,,
\end{array}\\
\chi_0^{(d)}(x,y) \stackrel{\rm def}{=} f(x,y-d)\;,\quad
\phi_0^{(d)}(x,y) \stackrel{\rm def}{=} f(y,x-d)\,.
\end{array}
$$
and
$$
\chi_t^{(d)} \stackrel{\rm def}{=} \Psi_t^{\chi_0^{(d)},U^{(d)}}\;,\quad
\phi_t^{(d)} \stackrel{\rm def}{=} \Psi_t^{\phi_0^{(d)},U^{(d)}}\;,\quad
\Psi_t^{(d)} \stackrel{\rm def}{=} \Psi_t^{\Psi_0^{(d)},U^{(d)}}\,.
$$
Obviously, if $V\in\testspace D(\bbbr)$ and $f\in\testspace S(\bbbr^2)\,$,
$$
\partial_\lambda\left(\partial_t^6 \int_{\cal G}\left(
\modulus{\Psi^{(d)}_t(x,y)}^2 - \modulus{\chi^{(d)}_t(x,y)}^2 -
\modulus{\phi^{(d)}_t(x,y)}^2\right){\rm d}x{\rm d}y\right)_{|_{t=0}}
\stackscript{\longrightarrow}_{d \to \infty} 0
$$
holds for every region ${\cal G}\subset\bbbr^3\times\bbbr^3\,$.

\noindent
Moreover,
$$
\begin{array}[c]{l}
\displaystyle \lim_{d\to\infty} \partial_\lambda\left( \partial_t^6
\int_{\cal G} \modulus{\phi^{(d)}_t(x,y)}^2\,{\rm d}x{\rm
d}y\right)_{|t=0}\\[8mm]
=\displaystyle \lim_{d\to\infty} \partial_\lambda\left( \partial_t^6
\int_{\cal G} \modulus{\chi^{(d)}_t(x,y)}^2\,{\rm d}x{\rm
d}y\right)_{|t=0}\\[8mm]
= c^{f,U}_{\lambda} \stackrel{\rm def}{=}\displaystyle \partial_\lambda\left(
\partial_t^6 \int_{x\in\cal O} \modulus{\Psi_t^{f,U}(x,y)}^2 {\rm
d}x\,{\rm d}y\right)_{|t=0}
\end{array}
$$
holds for
$$
\region G=\left\{(x,y)\in\bbbr^2: x\in\region O\,\lor\,y\in\region
O\right\}\;,\quad \region O \mbox{ bounded}\;,\quad U(x,y)=V(y)\,.
$$
Therefore, under these conditions,
$$
\partial_\lambda\left(\partial_t^6 \int_{\cal G}\left(
\modulus{\Psi^{(d)}_t(x,y)}^2\,{\rm d}x\,{\rm d}y\right)
\right)_{|_{t=0}} \stackscript{\longrightarrow}_{d \to \infty} =
c^{f,U}_\lambda\,.
$$
Since, as shown in Section \ref{section:bbm}, $c^{f,U}_{\lambda}$
can be arranged to be nonzero we conclude:
\begin{quote}
The postulate of symmetry or antisymmetry of the wave function w.r.t.~to
exchange of particles does not prevent essential nonlocality.
\end{quote}

\section*{Acknowledgments}

I am grateful to M.~Czachor, H.-D.~Doebner, G.A.~Goldin, and R.~Werner
for stimulating discussions.

\section*{Appendix: Maple V (Release 4) Session}

\scriptsize
\begin{verbatim}
PROCEDURES:
> del := proc(f)
>    global x,y,t;
>    option operator;
>    unapply( diff(f(x,y,t), x$2) + diff(f(x,y,t), y$2), (x,y,t));
> end:
> pot := proc(f)
>    global x,y,t;
>    option operator;
>    unapply(V(y) * f(x,y,t),(x,y,t));
> end:
> Idot := proc(f)
>    global x,y,t;
>    option operator;
>    unapply(simplify(subs(diff(P(x,y,t),t)=(-del(P)(x,y,t)+ pot(P)(x,y,t)),
  diff(PB(x,y,t),t)=(del(PB)(x,y,t)-pot(PB)(x,y,t)), diff(f(x,y,t), t))), (x,y,t));
> end:
EVALUATION:
> term0 := (x,y,t) -> 2 * diff(ln( PB(x,y,t) * P(x,y,t)), x) * PB(x,y,t) * 
  diff(P(x,y,t), x) + diff(ln( PB(x,y,t) * P(x,y,t)), x$2) * PB(x,y,t) * P(x,y,t):
> term1 := (x,y,t) -> simplify(Idot(term0)(x,y,t)):
> term2 := (x,y,t) -> simplify(Idot(term1)(x,y,t)):
> term3 := (x,y,t) -> simplify(Idot(term2)(x,y,t)):
SPECIAL CASE:
> spec := proc(f)
>     global x,y,t;
>     option operator;
>     unapply( subs(V(y)=lambda * y^2, P(x,y,t)=exp(-x^2 -y^2 -x*y),
  PB(x,y,t)=exp(-x^2 -y^2 -x*y), f(x,y,t)), (x,y,t));
> end:
RESULT:
> int(int(simplify(diff(spec(term3)(x,y,t),lambda)), x=-infinity..infinity),
  y=-infinity..infinity);

                                       1/2
                            - 32/3 Pi 3

> int(int(exp(-2*x^2 -2*y^2 -2*x*y), x=-infinity..infinity),
  y=-infinity..infinity);

                                     1/2
                             1/3 Pi 3
\end{verbatim}


\normalsize

\begin{thebibliography}{99}

\bibitem{Ashtekar}A.~Ashtekar and T.A.~Schilling, ``Geometrical
Formulation of Quantum Mechanics'', gr-qc/9706069.

\bibitem{Bona}P.~B\'ona, ``Quantum mechanics with mean-field
backgrounds'', Comenius University, Bratislava,
Phys.~Prepr.~Nr.~Ph10-91.

\bibitem{dg1}H.-D.~Doebner and G.A.~Goldin, ``On a general nonlinear
Schr\"odinger equation admitting diffusion currents'', \Journal{{\em
Phys.~Lett.} A}{162}{397--401}{1992}.

\bibitem{dg2}H.-D.~Doebner and G.A.~Goldin, ``Properties of nonlinear
Schr\"odinger equations associated with diffeomorphism group
representations'', \Journal{{\em J.~Phys.} A}{27}{1771--1780}{1994}.

\bibitem{Haag-Bannier}R.~Haag and U.~Bannier, ``Comments on {M}ielnik's
generalized (non linear) quantum mechanics'', \Journal{{\em
Commun.~Math.~Phys.}}{60}{1--6}{1978}.

\bibitem{Kibble}T.W.B.~Kibble, ``Relativistic models of nonlinear
quantum mechanics'', \Journal{{\em
Commun.~Math.~Phys.}}{64}{73--82}{1978}.

\bibitem{Kostin}M.D.~Kostin, ``On the Schr\"odinger-Langevin equation'',
\Journal{{\em J.~of Chem.~Phys.}}{57}{3589--3591}{1972}.

\bibitem{Mielnik}B.~Mielnik, ``Generalized quantum mechanics'',
\Journal{{\em Commun.~Math.~Phys.}}{37}{221--256}{1974}.

\bibitem{Sabatier}G.~Auberson and
P.C.~Sabatier, ``On a class of Homogeneous nonlinear Schr\"odinger
equations'', \Journal{{\em J.~Math.Phys.}}{35}{4028--4040}{1994}. 

\bibitem{Weinberg}S. Weinberg, ``Testing quantum mechanics'',
\Journal{{\em Annals of Physics} (NY)}{194}{336--386}{1989}.

\bibitem{tests}A.~Shim<ony, \Journal{{\em Phys.~Rev.}
A}{20}{394--396}{1979}.

C.G.~Shull, D.K.~Atwood, J.~Arthur, and M.A.~Horne:
\Journal{\PRL}{44}{765--768}{1980}.
                                
R.~Gaehler, A.G.~Klein, and A.~Zeilinger: \Journal{{\em Phys.~Rev.}
A}{23}{1611--1617}{1981}.

\bibitem{gis1}N. Gisin, ``Relevant and Irrelevant Nonlinear
Schr\"odinger Equations'', in {\em Nonlinear, Deformed and Irreversible
Quantum Systems}, eds. H.-D.~Doebner, V.K.~Dobrev, and P.~Nattermann
(World Scientific, Singapore, 1995), pp.~109--124.

\bibitem{gis2}N. Gisin, \Journal{\PRL}{53}{1776}{1984}.

%\bibitem{gis3}

N. Gisin, \Journal{\em Helv.~Phys.~Acta}{62}{363--371}{1989}.

N. Gisin, \Journal{{\em Phys.~Lett.} A}{143}{1--2}{1990}.

\bibitem{cl-sh}J.~F. Clauser and A. Shimony, \Journal{{\em Rep. Progr.
Phys.}}{41}{1881}{1978}.

\bibitem{czach1}M.~Czachor, %``Nonlocal-looking equations can make nonlinear quantum dynamics local'',
\Journal{{\em Phys.~Rev.} A}{57}{4122--4129}{1998}.

M.~Czachor, %``Complete positivity of nonlinear evolution: A case study'', 
\Journal{{\em Phys.~Rev.} A}{58}{128--134}{1998}.

%\Journal{{\em Phys.~Lett.} A}{225}{1}{1997}.



\bibitem{pol}J.~Polchinski, \Journal{\PRL}{66}{397--400}{1991}.

\bibitem{lu1}W.~L\"ucke, ``Nonlinear Schr\"odinger dynamics and
nonlinear observables'', in {\em Nonlinear, Deformed and Irreversible
Quantum Systems}, eds. H.-D.~Doebner, V.K.~Dobrev, and P.~Nattermann
(World Scientific, Singapore, 1995), pp.~140--154.

\bibitem{nlgt}H.-D.~Doebner, G.A.~Goldin, and P.~Nattermann, ``Gauge
Transformations in Quantum Mechanics and the Unification of Nonlinear
Schr\"odinger Equations'', quant-ph/9709036.

\bibitem{lu2}W.~L\"ucke, {\em Gisin Nonlocality of the Doebner-Goldin
2-Particle Equation\/}, quant-ph/9710033; not to be published. 

\bibitem{werner1}R.~Werner, June '97, unpublished\,.

\bibitem{NaDip}P.~Nattermann, ``Struktur und Eigenschaften einer Familie
nichtlinearer Schr\"odingergleichungen'', Diploma thesis, TU-Clausthal,
1993.

\bibitem{werner2}R.~Werner, March '98, unpublished\,.

\bibitem{nelson} E.~Nelson, \Journal{\em LNP}{262}{438--469}{1986}.

\bibitem{bbm}I.~Bialynicki-Birula and J.~Mycielski, \Journal{\em
Ann.~Phys.}{100}{62--93}{1976}.

\bibitem{kostal}H.~Ko\v s\v t\'al, \Journal{Acta
polytechnica -- Prag IV}{2}{123--130}{1973}.

\bibitem{luna}W.L.~and P.~Nattermann, ``Nonlinear Quantum Mechanics and
Locality'', in {\em Symmetry in Science X}, eds. B.~Gruber and M.~Ramek
(Plenum Press, 1998), pp.~197--205.

\end{thebibliography}
\end{document}